\documentclass[12pt]{article}
\usepackage{natbib}
\usepackage{graphicx}
\usepackage{epsfig}
\usepackage{amssymb}

\begin{document}

\baselineskip24pt

\centerline{\bf  Constraints on the Source of Lunar Cataclysm Impactors} 

\bigskip
\centerline{Matija \' Cuk$^1$, Brett J. Gladman$^2$, Sarah T. Stewart$^1$}

\bigskip

\centerline{$^1$Department of Earth and Planetary Sciences, Harvard University}
\centerline{20 Oxford Street, Cambridge, Massachusetts 02138, USA} 

\centerline{$^2$Department of Physics and Astronomy, University of British Columbia}
\centerline{6224 Agricultural Road, Vancouver, BC V6T 1Z1, Canada}

\bigskip

\centerline{E-mail: cuk@eps.harvard.edu}

\vspace{24pt}
\centerline{Re-submitted to Icarus}
\centerline{November 5$^{\rm th}$ 2009.}

\vspace{24pt}

\centerline{Manuscript Pages: 27}

\centerline{Figures: 2}

\centerline{Tables: 0}
\newpage

Proposed Running Head: Lunar Cataclysm Impactors

\vspace{48pt}

Editorial Correspondence to:

Matija \' Cuk

20 Oxford St

Department of Earth and Planetary Sciences

Harvard University 

Cambridge, MA 02138

Phone: 617-496-7712

Fax: 617-496-6958

E-mail: cuk@eps.harvard.edu

\newpage

\noindent {ABSTRACT: Multiple impact basins formed on the Moon about 3.8 Gyr ago in what is known as the lunar cataclysm or late heavy bombardment. Many workers currently interpret the lunar cataclysm as an impact spike primarily caused by main-belt asteroids destabilized by delayed planetary migration. We show that morphologically fresh (class 1) craters on the lunar highlands were mostly formed during the brief tail of the cataclysm, as they have absolute crater number density similar to that of the Orientale basin and ejecta blanket. The connection between class 1 craters and the cataclysm is supported by the similarity of their size-frequency distribution to that of stratigraphically-identified Imbrian craters. Majority of lunar craters younger than the Imbrium basin (including class 1 craters) thus record the size-frequency distribution of the lunar cataclysm impactors. This distribution is much steeper than that of main-belt asteroids. We argue that the projectiles bombarding the Moon at the time of the cataclysm could not have been main-belt asteroids ejected by purely gravitational means.}  
\bigskip


\newpage 

\vspace{24pt}

\centerline{\bf 1. Introduction}

The lunar cataclysm was proposed \citep{ter74} on the basis of the ages of numerous impact-melt rocks, collected by the Apollo astronauts on the nearside, clustering in the narrow interval of 3.8-4 Gyr ago (Gya). The lack of older impact melts is most often interpreted in two very different ways: either an intense bombardment reset geochemical clocks to the formation of the last few basins, recording the tail-end of planetary accretion \citep[``stonewall'';][]{har75}; or there was little bombardment between planetary accretion and an impact spike at about 3.9 Gya \citep{ryd90}. The latter view came to dominate, but the controversy has endured due to the lack of post-Apollo lunar samples. While dates for the Serenitatis basin formation consistent with an impact spike have been produced by more recent dating of some Apollo 17 samples \citep{dal96}, other studies seem to support prolonged basin formation, or at least a pre-spike formation of Nectaris  \citep{war03}. The data from impact dating of meteorites (originating in the asteroid belt) ``differ significantly'' from those based on Apollo samples \citep[][ see also Hartmann, 2003]{ccg07} making some scientists question the Solar System-wide nature of the cataclysm \citep{ryd90}. 

There has been a recent revival in the study of the lunar cataclysm, motivated in part by advances in planetary dynamics. The Nice model of planetary migration may explain giant planet eccentricities and inclinations \citep{tsi05}, as well as the origin of Jupiter Trojans \citep{mor05}, Hildas, and asteroid belt comets \citep{lev09}. The main feature of the model is a past divergent migration of Jupiter and Saturn through their mutual 1:2 mean-motion resonance, triggering a relatively short interval during which the outer planet system expanded into a previously stable outer comet belt. While the main events in the Nice model take place over only a few Myr, it is possible to delay the onset of the rapid migration phase for hundreds of Myr, opening the possibility that the lunar cataclysm was a consequence of this delayed planetary migration \citep{lev01, gom05}. The impactors are initially drawn from the outer Solar System, but about 100 Myr after the initial resonance crossing, main-belt asteroids ejected by the $\nu_6$ secular resonance \citep[which sweeps the asteroid belt as Saturn migrates;][]{min09} begin to dominate in the inner Solar System.

In an influential paper often taken to support the connection between the Nice model and the lunar cataclysm, \citet{str05} argue that the crater size-frequency distribution on the lunar highlands matches the production function of the asteroid belt (using old crater data and new asteroid observations). The implication is that the ancient asteroid belt (taken to have the same size distribution as today) was destabilized by a purely gravitational mechanism (such as a sweeping secular resonance). Younger terrains (and morphologically young craters) preserve a different size distribution, which \citet{str05} argue is similar to the present near-Earth asteroids (NEAs). NEAs also originate in the asteroid belt but are ejected by the size-dependent Yarkovsky effect \citep{mor03}, leading to a relative overabundance of smaller bodies among NEAs compared to the main-belt asteroids (MBAs).

 In this work, we critically examine the lunar crater record of the final stages of the cataclysm.

\bigskip

\centerline{\bf 2. Defining the Cataclysm}

\bigskip

In a recent review, \citet{ccg07} concluded that one can say with certainty only that the lunar cataclysm ended relatively abruptly 3.8 Gya and that it is impossible to say when it started based on the available data. While we share this view, we believe that dynamical modeling can help constrain the beginning and the nature of the cataclysm. \citet{bot07} modeled the decay of a primordial high-inclination NEA-like population and found that after about 100 Myr of fast depletion (with half-life of 15 Myr) the leftover population decays with a half-life of about 80 Myr. The conclusions of \citet{bot07} are very close to those of other direct numerical simulations of small body depletion in the inner Solar System \citep{gla00, mor01}. For reasonable initial populations at 4.5~Gya, this depletion leaves too few impactors to plausibly produce the indisputably young Imbrium and Orientale basins in the 3.8-3.9 Gya window. More numerous initial populations are efficiently ground down by mutual collisions and cannot survive for 600 Myr \citep{bot07}. 

Given that the dating of all basins except Imbrium is disputed, we use the term ``lunar cataclysm'' only for the late and temporally-close formation of the Imbrium and Orientale basins. According to \citet{bot07}, even such a ``minimalist'' lunar cataclysm must have been an event separate from planet formation (as primordial planetesimals were all but eliminated by this time). 

Using this definition, only those lunar surface units that are indisputably close in age to the Imbrium and Orientale basins can be used to study the cataclysm. This definition is pragmatic and does not reflect any opinion on the nature of the cataclysm. It is likely that at least some of the pre-Imbrium basins also formed at this time, but the arguments presented here do not depend on it.

We will not use any cratering data from Mars or Mercury to study the crater-size distribution of lunar cataclysm impactors. While Mars may have also suffered intense bombardment at 3.9 Gya \citep{ash96}, no terrain on Mars has an absolute date associated with it. Relative dating by crater counts is sometimes converted to absolute dates by assuming that the lunar and martian bombardment histories were the same, but any dates derived this way cannot be considered independent. One can compare the size-frequency distributions of martian and lunar craters to probe a possible change in impactor populations \citep{str05}, but the absolute chronology will have to be calibrated using only lunar samples until we have martian surface samples collected in context.

\bigskip

\centerline{\bf 3. Imbrian Impact Chronology}

\bigskip

Early lunar history is divided into three systems \citep{wil87}: {\it pre-Nectarian}, consisting of surfaces older than the Nectaris basin, {\it Nectarian}, spanning the interval of uncertain duration between the Nectaris and Imbrium impacts, and the {\it Imbrian} system, postdating the formation of the Imbrium basin and lasting until about 3.2 Gya. The Imbrium and the younger Orientale basins are considered lower Imbrian terrains, while the numerous maria\footnote{A {\it basin} is an impact crater larger than about 300~km; {\it mare} is an area of basaltic lava flows, usually darker than the surrounding terrain. When a mare fills a basin, the mare must be younger than the basin itself.} formed subsequently by basaltic volcanism are considered upper Imbrian units.

\begin{figure*}[h]
\includegraphics[scale=.6, angle=0]{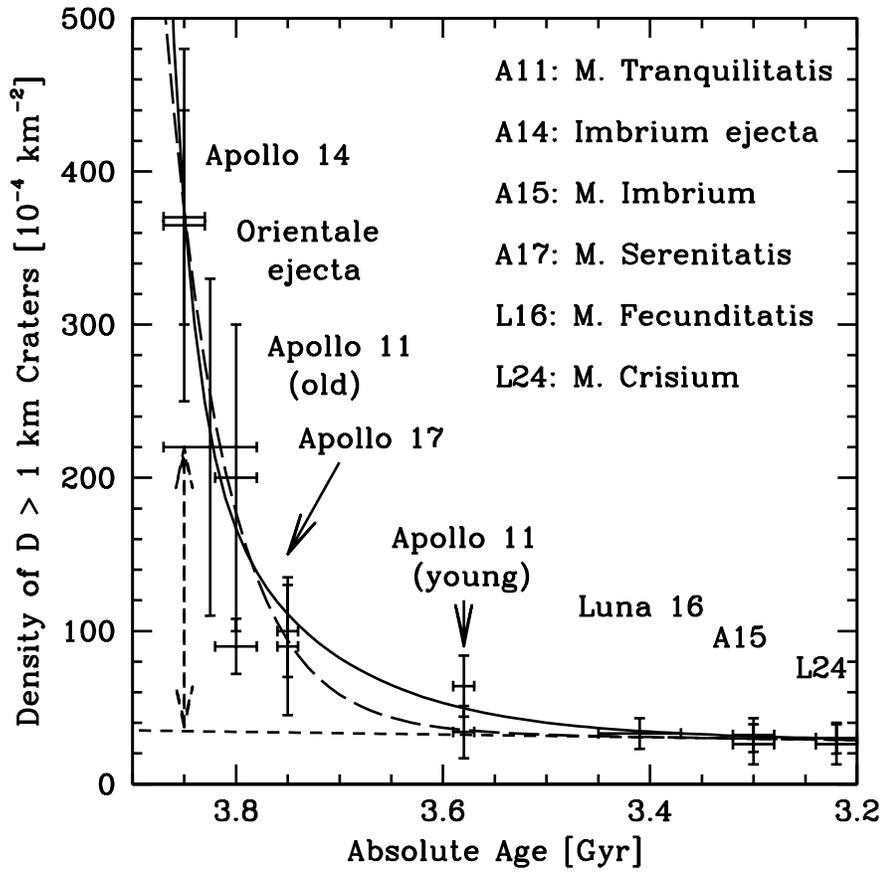}  
\caption{\' Cuk, Gladman and Stewart, Lunar Cataclysm Impactors} 
\label{tail}
\end{figure*}

Figure \ref{tail} plots the crater densities on eight Imbrian units versus the radiometric age of the associated samples (in the case of the unsampled Orientale basin, the age error bar is defined by neighboring Imbrium and Mare Tranquilitatis samples). Absolute ages are derived from dating of returned lunar samples. All ages and crater densities (together with their error bars) are taken from \citet{sto00}, with (probably overestimated) crater count error bars of 50$\%$ assumed where no uncertainty is provided. 

What can we conclude about the Imbrian chronology? The data are consistent with a simple depletion of an Earth-crossing impactor population which is not being replenished \citep[][ solid line in Fig.\ \ref{tail}]{bot07}. An instantaneous injection of an impactor population into the inner Solar System not long before the Imbrium impact is one plausible interpretation. Many other scenarios are allowed, but there are two that can be excluded. A purely geocentric cataclysm with no impactors escaping into heliocentric orbit would be geologically instantaneous and can be ruled out by the elevated crater counts on the old parts of Mare Tranquilitatis and Mare Serenitatis. A very slowly-declining continuing supply of new Earth-crossers (with a half-life $\ge $100 Myr) would produce a cataclysm tail even longer than that shown in Fig.\ \ref{tail}, which would be irreconcilable with the crater populations on both the basins and the maria. The simplest explanation is that the remaining impactors in near-Earth space at the time of Orientale's formation were simply eliminated with the natural dynamical half-life, providing the tail of the cataclysm.

The most important implication from Fig. \ref{tail} is that the majority (66\%-90\%) of craters on the Orientale basin ejecta blanket were produced by the impactors belonging to the tail of the Lunar Cataclysm, rather than the more uniform post-3.5 Gya component. The background cratering by NEAs (short-dashed straight line) should have added less than about $40 \times 10^{-4} \ {\rm km}^{-2}$ craters to the Orientale ejecta blanket, which has a $D>$1-km crater density above $200 \times 10^{-4} \ {\rm km}^{-2}$ (vertical dashed arrow). 

Despite poor statistics of $D>1$~km craters on Orientale's ejecta (this size range was used in Fig. 1 as it provides the best data for maria), the association of most post-Orientale craters with the cataclysm is well established, with independent data \citep[][their Table 8.3.2]{har81} showing that Imbrium and Orientale ejecta blankets respectively have 3 and 2.5 times the crater density (using $D>2.8$~km) of the reference ``average mare'' (an abstract concept, used there as a unit of crater density). Late Imbrian maria like Mare Imbrium and Mare Crisium (which Fig.\ \ref{tail} suggests formed after even the tail end of the cataclysm finished) according to \citet{har81} have about half of the ``average mare'' crater density, supporting the view that about 80\% of the craters on the Orientale basin and ejecta blanket were formed during the cataclysm.

Our use of the \citet{bot07} results to fit to model the data plotted in Fig.\ref{tail} could possibly lead to some confusion concerning the end of the lunar cataclysm. Despite the fact that \citet{bot07} curve is a combination of two exponential functions, it still describes the depletion of a single impactor population, namely the high-inclination NEAs. This is because the initial flood of impactors into Earth-crossing space fills both rapidly-declining and longer-lived meta-stable orbital regions. Therefore the ``knee'' between the two components (at 3.8~Gya) does not imply a change in the nature of impactors or an end to the cataclysm. If we accept the simplest interpretation that there are two components that have contributed to the crater densities in Fig.\ \ref{tail}, the background and the Cataclysm, the background steady-state NEA cratering rate alone can explain only the crater densities for 3.6~Gya and younger terrains. 

\bigskip

\centerline{\bf 4. The Lunar Cataclysm Crater Size-Frequency Distribution}

\bigskip

Depending on each particular hypothesis for cataclysm event, there are different ways of deriving the size-distribution of cataclysm impactors from the lunar crater record. Some researchers think that all of the Moon was resurfaced by impacts in the 3.9-4 Gya time range (this lack of older ages is essential to the stonewall hypothesis, but not required by the spike scenario). Assuming total resurfacing during the cataclysm, \citet{str05} suggest that all heavily cratered lunar surfaces preserve an impactor size distribution very similar to that in today's asteroid belt. To reach this conclusion, it is necessary to assume that all of the lunar highlands were resurfaced (i.e. their crater retention ages were reset) in the cataclysmic impact spike, despite a complete lack of absolute dates for heavily cratered terrains. Here we offer a different approach that uses only those portions of the Moon which have been either dated or bracketed by absolute dates and stratigraphy. In particular, we focus on the crater densities on Imbrian terrains, which are not close to crater saturation and must have been subject to the impactors from the tail of the cataclysm (Fig.\ \ref{tail}).

\begin{figure*}[h]
\includegraphics[scale=.6]{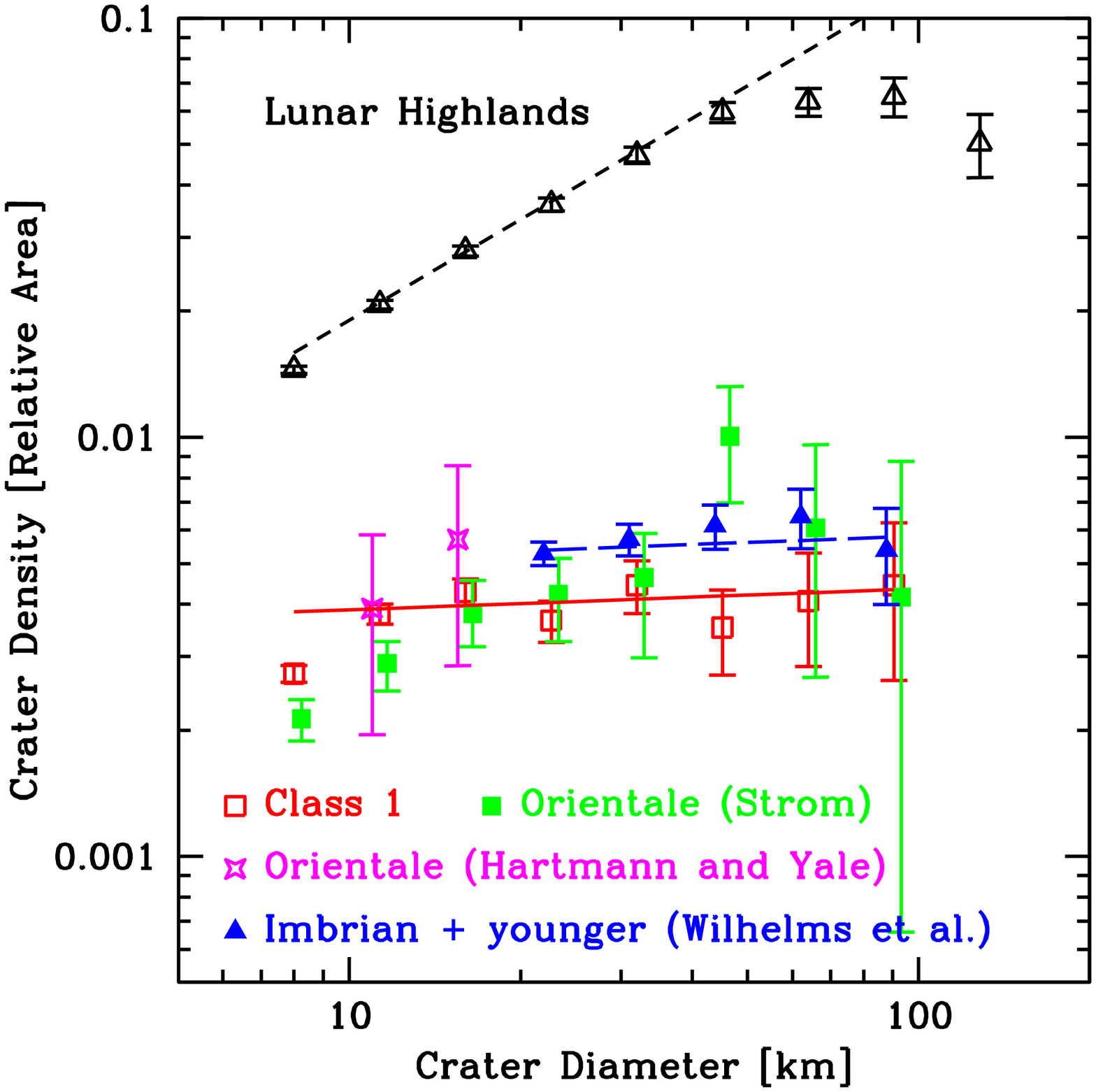}
\caption{\' Cuk, Gladman and Stewart, Lunar Cataclysm Impactors}
\label{orientale}
\end{figure*}

 Since Orientale was the last large impact basin, its ejecta blanket should preserve a pristine record of the subsequent impactor flux (free of any pollution from other basin secondaries or the crater erasure sometimes proposed for older terrains). In order to isolate these upper Imbrian impactors belonging to the tail of the cataclysm, \citet{str77} counted craters on the Orientale ejecta blanket; the results are plotted in Figure \ref{orientale} (solid squares). Smaller-scale counts of 10-km diameter craters on the Orientale ejecta blanket by \citet{har68} agree (stars). The much-denser total highland crater population is plotted (open triangles) for comparison. \citet{str77} used the similar shape of the two size-frequency distributions to conclude that craters on Orientale's ejecta blanket have the same size distribution as those on the lunar highlands. Later, \citet{str05} connected the latter to the production function of the main belt asteroids (with the differential logarithmic-bin slope of about -1.3) and  the different size-distribution on younger terrains (with the differential log-bin slope of about -2) was attributed to impacts by NEAs.\footnote{Some researchers find a single crater size-frequency distribution on all lunar terrains \citep{neu01}.}

However, crater counting on younger terrains is not the only way of isolating young craters. \citet{str05} present the size distribution of morphologically-fresh class 1 highland craters (open squares in Fig. \ref{orientale}) as evidence of a late (post-lunar cataclysm) impactor size distribution similar to that of NEAs. Surprisingly, the {\it absolute surface densities} of 10~to~100-km diameter class 1 craters are (given the uncertainties) the same as those on the Orientale ejecta blanket. If we accept the reasonable assumption, as \citet{str05} do, that the class 1 craters are the youngest craters in relative age, one must conclude that the first class 1 craters date back to about the time of the Orientale impact. Since the Orientale basin formation was followed by a fast decay in impact rates as Earth-crossing space was dynamically cleared of impactors (Fig. \ref{tail}), most craters on the Orientale ejecta blanket must have formed during the tail end of the cataclysm. As class 1 craters and those on the Orientale basin and ejecta blanket have the same absolute crater density, most class 1 craters must have also formed during the period of rapid fall-off in bombardment illustrated by Fig. \ref{tail}.

Unless one accepts an extremely non-uniform past distribution of lunar impacts, the equal areal density of these two populations requires that they come from the same projectile population, and thus share the same impactor size distribution. While class 1 crater-count statistics are better than those for Orientale's ejecta blanket, there is always a possibility that there was some bias in morphological classification. In order to check if class 1 craters can be trusted as an unbiased record of late Imbrian impactors, in Fig.\ \ref{orientale} we also present data for Imbrian plus post-Imbrian craters larger than 20~km in diameter counted by \citet{wil78}. \citet{wil78} primarily used stratigraphy to assign craters to different lunar systems, so their approach should be immune to potential biases of a strictly-morphological classification. Note that many of these Imbrian craters come from terrains resurfaced by the Imbrium impact, and therefore their number density is unsurprisingly slightly higher than that on Orientale's ejecta blanket. On the size interval they cover, \citet{wil78} counts are consistent with a ``flat'' log-bin differential exponent of about -2, appearing unsurprisingly as a ``scaled up'' version of the class 1 distribution.  This is consistent with the class 1 and Imbrian craters reflecting the same impactor population, the majority of which struck {\it during the tail of the lunar cataclysm} rather than during the subsequent 3.5 Gyr.

It is important to note that the connection between class 1 craters and the lunar cataclysm does not rest on the size-frequency distribution of craters superposed on the Orientale basin, but rather on their absolute density. The error bars on the Orientale counts alone are too large to unequivocally match its size-distribution to either class 1/Imbrian or the highland/asteroid curves (Orientale counts are consistent with either within the uncertainties). However, the overlapping crater densities of the Orientale basin and class 1 craters imply that the two samples {\it must come from the same epoch} and were formed by the same impactor population (which is also consistent with the Imbrian sample). With this fact established, we can use class 1 craters to make inferences about the lunar cataclysm impactors.

\bigskip

\centerline{\bf 5. Implications for the Source of Imbrian Impactors}

\bigskip

The size-frequency distribution we find for the lunar cataclysm impactors (slope of -1.9 or -2) is not very different from that of current NEAs \citep[-1.75 to - 1.8, ][]{mor03}. However, these must be two distinct dynamical populations because the gradual thermal Yarkovsky effect that produces NEAs could not have conceivably produced a bombardment spike. While this similarity in slope makes distinguishing between these two populations hard (especially on terrains with poor crater statistics), there is no a priori reason why two populations of dissimilar origin cannot have similar size-distribution slopes. Saying that class 1 craters are obviously caused by NEAs without any crater-density and chronological arguments is insufficient.

 While the size distribution of trans-neptunian objects in the relevant size range is unknown, comets alone are unlikely to have produced the lunar cataclysm. \citet{gom05} show that tens of Earth masses of trans-neptunian objects are needed to produce a lunar cataclysm (due to low impact probabilities). Depletion of such a massive planetesimal disk inevitably leads to large-scale planetary migration and $\nu_6$ secular resonance sweeping the asteroid belt, that is, the events proposed in the Nice model \citep{lev01, min09}. \citet{gom05} also find that escaped asteroids remain in Earth-crossing orbits longer than the destabilized comets, so the bombardment tail in the Nice-model-type cataclysm should be dominated by asteroids even if both populations were destabilized simultaneously. Therefore it is hard to imagine a scenario in which the tail of the cataclysm would consist primarily of comets derived from the trans-neptunian region.

Given that the lunar cataclysm was produced by a population having a differential size-frequency distribution described by a power law with an index of -1.9 or -2 rather than -1.2 or -1.3 \citep[like the modern asteroid belt;][]{ive01}, theoretical models producing the lunar cataclysm by gravitational ejection of main-belt asteroids are seriously challenged. These scenarios include the Nice model\footnote{It is important to note that our results do not imply that the Nice model-type planetary migration did not happen, but that it occurred long before the lunar cataclysm and that the two are unrelated.} and the Planet V \citep{cha07} hypothesis. The Planet V scenario proposes that an additional planet managed to survive for almost 1 Gyr on an orbit between Mars and the asteroid belt, destabilizing the belt during its escape. 

Our conclusion about the incompatibility of the lunar cataclysm impactors with the main-belt asteroids holds only if the main-belt size distribution at the time of release of the cataclysm impactors was not significantly different from what it is now. The size distribution that a collisionally-evolving population reaches depends primarily on impact speeds and the material properties of the bodies in question. Recent theoretical work  \citep{bot05} indicates that, if the dynamical state of the belt was not significantly changed at the time of the cataclysm, the main-belt size distribution would have already been established, and thus should have been the same as the present day at the time of the lunar cataclysm (in this case one can conclude the lunar cataclysm impactors do not match the asteroid belt). Alternately, perhaps before the cataclysm event the main asteroid belt was dynamically colder and had a different size distribution; in this scenario the event that releases the impactors heats the belt and thus starts a collisional degradation to the current size distribution. However, \citet{min08} show that if the asteroids had much less eccentricity than today, then when the $\nu_6$ resonance sweeps the belt (the most plausible mechanism to eject huge numbers of asteroids) the belt's destruction is far too efficient to leave the current belt behind. Furthermore, an exceptionally massive cataclysm originating from a dynamically cold asteroid belt is inconsistent with the cataclysm-linked crater size distribution being found on Imbrian but not most older terrains (Fig. \ref{orientale}). Further work is clearly needed, but the issue of a possible change in the main-belt size distribution does not affect our main conclusion that the cataclysm impactors do not match the current asteroid belt.

What are the alternatives to asteroidal impactors? It is difficult to imagine an additional small-body population surviving in great numbers until about 3.8 Gya and then becoming rapidly depleted. Given that Fig. 1 allows for the instantaneous injection of the lunar cataclysm impactor population into the inner Solar System, one alternative is a late disruption of a single large body (with $D\geq 500$~km if disrupted on an Earth-crossing orbit) in order to produce the required impactors \citep{wet75}. The disruption would likely have to be tidal \citep{asp06}, as a catastrophic collision involving {\it two} large inner solar system planetesimals leftover from accretion would have been much less likely than a late survival of one of them \citep{wet75}. Alternatively, a collision of an additional terrestrial planet with Mercury (or Venus) in which the planet is disrupted and re-accreted cannot be completely ruled out without samples or meteorites from those bodies. In any case, the trigger for the cataclysm would have been the dynamical destabilization of a long-lived large body. Some long-duration quasi-stable orbits are known to exist in various pockets of the inner Solar System \citep{tab00, cha07}, but it is not clear the these regions ever contained much mass. More recently, \citet{min09b} have suggested that as many as 30\% of large asteroids may become unstable more than 200 Myr after the final sculpting of the belt. Given that the asteroid belt contains several bodies of the size sufficient to cause the lunar cataclysm, this avenue of research appears promising, but more work is needed to assess the probability of a massive tidal disruption event.

\bigskip

\centerline{\bf 6. Conclusions}

\bigskip

Our conclusions are summarized as follows:

1. The correlation of lunar crater counts and radiometric dating of lunar samples support an impact spike at about 3.8 Gya \citep{ter74}.

2. Most of the craters on the Orientale basin and ejecta blanket formed during this impact spike (like the Orientale basin itself).

3. Morphologically-young class 1 craters all over the nearside highlands, which have an absolute density similar to that of craters on the Orientale basin and ejecta blanket, also formed during this bombardment spike. 


4. The size-frequency distribution of class 1 craters thus records the cataclysm impactors. This distribution is not the same as that of the current main asteroid belt.

5. The size-frequency distributions of craters assigned to the Imbrian period \citep{wil78}, which includes the Orientale basin, agree with that of class 1 craters but not with current main-belt asteroids.

6. Assuming that the asteroid belt had the same size-distribution 3.9 Gyr ago, asteroids ejected by sweeping secular resonances or scattered by a surviving protoplanet could not have caused the lunar cataclysm.

\vspace{48pt}

{\centerline{\bf ACKNOWLEDGMENTS}}

\bigskip

M\' C is a Daly Postdoctoral Fellow at Harvard University. We wish to thank Clark Chapman and Bill Hartmann for very helpful discussions, as well as Zo\" e Leinhardt and Laurel Senft for their insightful comments on an earlier draft.

\bibliographystyle{}

\newpage

{\bf FIGURE CAPTIONS}

\bigskip

{\bf Figure 1.} The tail of the lunar cataclysm. Number density of craters (larger than 1 km in diameter per 10,000~km$^2$) versus age for eight Imbrian terrains labeled on the plot. All data are taken from \citet{sto00}. Most of the terrains have more than one crater count reported in the literature, which we plot as separate points. The solid line plots a possible fit based on results of \citet{bot07}, with the transition between 15 and 80 Myr decay half-lives at 3.8 Gya, while the long dashed line has constant half-life of 40 Myr (both approximations include a background term linear in time). The straight short-dashed line shows contribution from a constant background bombardment which reaches zero density today (the slope is chosen to fit the last two Mare surfaces in the plot). The dashed arrows show the excess of craters on Orientale ejecta blanket formed during the tail of the lunar cataclysm.

\

{\bf Figure 2.} Class 1 craters have similar number density to those on Imbrian terrains. Points plot the number of craters in the diameter range D to $\sqrt{2}$D, multiplied by $\sqrt{2} \pi D^2 / 4$ and divided by the counting area. A population with a logarithmic-bin differential size-frequency exponent of -2 would appear as a horizontal line. Open squares represent morphologically fresh class 1 craters from nearside highlands \citep{str05}, while the solid squares and stars plot counts on the Orientale ejecta blanket by \citet{str77} and \citet{har68}, respectively (slightly shifted to the sides for clarity). For reference, we plot slopes of -1.95 (solid and long-dashed line) and -1.2 (short-dashed line) which are approximate matches for Class 1 and total highland \citep[open triangles; ][]{str05} crater size-frequency distributions. Solid triangles plot the density of Imbrian and post-Imbrian craters counted by \citet{wil78}. While having a higher density, the latter group is also consistent with the size-frequency distribution of class 1 craters (and fit the long-dashed line with a slope of -1.95) but not the total highland crater population.


\newpage

\end{document}